# Physics-Enforced Neural Ordinary Differential Equation for Chemical Kinetics Optimization in Reaction-Diffusion Systems

Feixue Cai[a], Hua Zhou[a], Zhuyin Ren[a],*

[a] *Institute for Aero Engine, Tsinghua University, Beijing 100084, China*
* *Corresponding author:* zhuyinren@tsinghua.edu.cn

**Abstract**

Calibrating chemical kinetics in a reaction-diffusion system is challenging because of complex dynamics governed by tightly coupled chemistry and transport, while experimental observations are often sparse and noisy. We propose a physics consistent diffusion-chemistry coupled neural ordinary differential equation (Diff-Chem Neural ODE) that embeds Arrhenius-structured reaction neurons into a fully differentiable streamline formulation and explicitly accounts for diffusion coupling. This design enables direct gradient-based analysis of kinetic parameters without sampling-based pretraining. We validate this method on burner-stabilized flat and stagnation reacting flows using mechanisms spanning different stiffness ranges. The proposed method reproduces species profiles with near-reference accuracy, whereas a pure chemistry Neural ODE that neglects diffusion coupling may misplace ignition and generate an incorrect thin reaction zone. Diff-Chem Neural ODE is more robust than pure chemistry Neural ODE and provides substantial speedups for gradient evaluation compared with fully discretized computations. In kinetics refinement, optimizing only a limited set of "primal" species reduces the loss by over 98% and simultaneously recovers unobserved variables, demonstrating physically consistent global control. Finally, tests with 1-20% noise in the objective show stable convergence without local overfitting, supporting its applicability under noisy measurements.

## 1. Introduction

Energy conversion and propulsion devices are governed by the coupled evolution of temperature and species fields under transport and chemical transformation [1-3]. Similar coupled transport-chemistry dynamics also arise in a broad class of reacting media beyond energy systems, including heterogeneous catalytic reactors in chemical processing [4,5], reaction-diffusion fronts in nano materials [6] and biological media [7] etc. In continuum form, the correspond dynamics are often represented as an integrated Reaction-Diffusion (R-D) system or Reaction-Diffusion-Convection (R-D-C) system in which transport processes interact with reactive process [1]: i) advection (flow transport); ii) molecular species diffusion; iii) heat conduction; iv) chemical source terms. These systems form a series of Partial Differential Equations (PDEs) structure, which implies that the state of an operating device is determined not only by intrinsic chemical time scales but also by transport-controlled access to, and residence on, specific regions of thermo-chemical phase space [4,8]. Consequently, the fidelity of reaction kinetics has direct implications for predicting and optimizing stability margins, conversion efficiency, and pollutant formation, particularly when temperature composition feedback and gradient-driven fluxes are significant.

The inclusion of diffusion/convection introduces nonlinear terms and strong coupling to the velocity field, which substantially increases analytical and numerical complexity, particularly in regimes with sharp gradients [9]. Thus, the transport and chemistry coupling system becomes decisive near operational limits, where small deviations in kinetics can lead to large shifts in stability boundaries [10-12]. As an example, in canonical burner-stabilized configurations, the dominant gradients are aligned with a principal direction and the system is frequently treated using quasi-one-dimensional descriptions [13]; this structure is valuable precisely because it isolates the coupled effects of chemistry and molecular transport (including differential diffusion and heat release) under well-controlled boundary conditions. Early measurements in hydrocarbon diffusion flames already emphasized the tight interplay between chemical pathways and transport processes [14]. The complementary burner-stabilized stagnation configuration introduces a controlled strain field and strong flow-chemistry interaction, enabling systematic interrogation of kinetics-transport coupling and, when relevant, extinction/near-extinction behavior [15]. These canonical examples highlight that, in realistic operating regimes, diffusion and convection can directly condition the reachable states and their stability, rather than acting as secondary perturbations.

Despite this, for energy conversion systems, kinetic mechanism calibration and optimization for hydrocarbon and related fuels has predominantly been performed in homogeneous reactor settings, where governing equations reduce to an Ordinary Differential Equation (ODE) system under perfect -mixing) assumptions [16,17]. While such optimization is effective for matching ignition delays and perfectly-stirred reactor data, it is structurally limited in representing transport-constrained trajectories and flow-reaction coupling that emerge in spatially inhomogeneous reactors. In particular, objectives defined solely on well-mixed ODE trajectories do not encode diffusion- and convection-induced constraints that become prominent in strained flames, wall-influenced reacting layers, or other near-limit regimes [18]. As a result, ODE-only optimization may under-sample or entirely omit the portions of phase space that are most relevant for stability and performance in practical devices. The above limitations are closely related to the theory of low-dimensional structure in reacting systems. For homogeneous kinetics, chemical-manifold methods exploit stiffness-induced attraction toward slow sets [19-23]. These methods characterize fast/slow subspaces and support systematic reduction of detailed chemistry [19,24]. However, when spatial transport is comparable to reaction in conditioning the evolution, the relevant attracting sets are those of the coupled PDE operator, and they generally do not coincide with the slow sets implied by the homogeneous ODE alone [25]. This motivates Reaction-Diffusion (R-D) manifold approaches that incorporate transport effects into reduced state-space descriptions [25-27].

At the same time, mechanism refinement has also long incorporated R-D-C informed observables, including laminar flame velocities, flame-structure quantities, and extinction-related metrics in canonical 1D strained-flame configurations [28,29]. The primary practical limitation of such R-D-C driven calibration is often not the absence of transport physics in the objective, but the computational burden associated with repeatedly evaluating transport-chemistry coupled simulations over wide condition spaces and high-dimensional parameter sets. This burden becomes particularly acute when detailed or skeletal mechanisms with $\mathcal{O}(10\sim10^3)$ reactions and $\mathcal{O}(10^2\sim10^4)$ species are optimized [30], as each forward evaluation requires solving a stiff, coupled transport-reaction boundary value problem, and the number of evaluations required by uncertainty quantification, or Bayesian inference workflows can be substantial [29,31-33].

Recent advances in scientific machine learning provide additional tools to operationalize this requirement [33,34]. Neural Ordinary Differential Equations (Neural ODEs) [35] parameterize the time derivative of a state by a learnable function and compute trajectories using ODE solvers, enabling end-to-end training through differentiable integration. For reacting systems, this paradigm offers a natural mechanism for gradient-based parameter learning at the level of continuous-time dynamics, and has been extended to address stiffness and multiscale kinetics through algorithmic and architectural modifications (e.g., stiff Neural ODE formulations and stabilized training strategies) [36]. Moreover, domain-structured Neural ODE variants can hard-encode chemical interpretability: the Chemical Reaction Neural Network (CRNN) architecture embeds the law of mass action and Arrhenius kinetics directly into the network so that weights/biases correspond to stoichiometry and kinetic parameters [37-39]. And the chemical reactor nets can be modeled through Neural ODE and non-ideal reactors [40,41]. More generally, universal differential equation formulations provide a unifying perspective in which mechanistic models are augmented with learnable components while retaining differentiable simulation for parameter estimation and model discovery [42]. In parallel, Physics-Informed Neural Networks (PINNs) embed PDE residuals into training objectives and have been used to infer solutions and parameters of nonlinear PDE systems from sparse data, thereby enabling inverse problems under governing-equation constraints [43]. In kinetics contexts, representative "chemistry-aware PINN" implementations explicitly incorporate mass-action and conservation laws into the PINN loss to reduce nonphysical errors and data dependence [44,45]. Complementarily, recent work demonstrates that enforcing conservation laws in neural surrogates can be done not only via soft penalties but also via plug-and-play correction constraint strategies, substantially improving stability in long-horizon reactive evolution [46].

However, existing Neural ODE- and PINN-based paradigms do not, in their standard formulations, directly address the specific challenge posed by R-D native kinetics optimization. Most Neural ODE applications to kinetics learning are trained on effectively 0D (well-mixed) trajectories, so the learned continuous-time vector field is informed primarily by homogeneous evolution and is not systematically exposed to the transport-admissible pathways and constraints induced by diffusion and convection in spatially inhomogeneous reactors [37]. Conversely, PINNs enforce PDE consistency by embedding residuals into a global training objective, but for stiff, multiscale reacting systems this monolithic enforcement can be difficult to optimize [47] and may not naturally preserve reaction-level interpretability (e.g., pathway attribution and identifiable kinetic parameters) that is essential for mechanism refinement. These observations suggest a gap between i) data-driven continuous-time learning that is efficient and differentiable but often transport-agnostic, and ii) PDE-constrained learning that is transport-consistent but can be computationally and algorithmically challenging for stiff detailed chemistry and may obscure mechanism structure. Therefore, a transport-consistent learning formulation for kinetics optimization should: i) retain reaction-level structure, ii) remain differentiable for gradient-based optimization under stiffness, and iii) expose the learning process to transport-constrained trajectories that characterize real reactors.

Therefore, we propose a physics-enforced diffusion-chemistry coupled neural ordinary differential equation (Diff-Chem Neural ODE) for efficient, transport-consistent kinetics analysis and optimization in R-D systems. Section 2 reformulates the steady reacting-flow PDEs into a residence-time/streamline ODE for differentiable integration, and presents the Diff-Chem Neural ODE with Arrhenius-structured reaction neurons and diffusion modeled as explicit forcing. Section 3 describes the test cases, validates against reference solutions, and evaluates robustness and efficiency under perturbed initial conditions, noisy measurements, and runtime/gradient benchmarks. We then demonstrate kinetics refinement using objectives defined on a limited set of measurable species. Conclusions are given in Section 4.

## 2. Theoretical formulations

*2.1. Governing equations for reaction-diffusion-convection systems*

As mentioned above, the transport due to molecular diffusion and velocity will introduce diffusion and convection effects into the reaction process. Considering a typical reacting flow with $N_S$ species and $N_R$ reactions, it can be described by series of PDE system including mass, momentum, energy and species conservations (where the body forces are neglected for common cases). In general, the governing equations can be expressed as:

$$\frac{\partial \rho}{\partial t} + \nabla \cdot (\rho \boldsymbol{U}) = 0 \tag{1}$$

$$\frac{\partial (\rho \boldsymbol{U})}{\partial t} + \nabla(\rho \boldsymbol{U} \otimes \boldsymbol{U}) = -\nabla P + \nabla \cdot \left[\mu \nabla \boldsymbol{U} + \mu (\nabla \boldsymbol{U})^{\top} - \frac{2}{3}\mu(\nabla \cdot \boldsymbol{U})\mathbf{I}\right] \tag{2}$$

$$\frac{\partial (\rho Y_i)}{\partial t} + \nabla \cdot (\rho \boldsymbol{U} Y_i) = -\nabla \cdot \boldsymbol{J}_i + M_i \dot{\omega}_i, \quad \sum_{i=1}^{N_S} Y_i = 1 \tag{3}$$

$$\frac{\partial \rho c_p T}{\partial t} + \nabla \cdot (\rho c_p \boldsymbol{U} T) - \frac{\partial P}{\partial t} = \nabla \cdot (\lambda \nabla T) - \nabla \cdot \left(\sum_{i=1}^{N_S} h_i \boldsymbol{J}_i\right) - \sum_{i=1}^{N_S} M_i h_i \dot{\omega}_i \tag{4}$$

Here, $\nabla$ is the variable's gradients to Eulerian locations and $\otimes$ the tensor product. The $\rho$ and $\boldsymbol{U}$ are the density of mixture fluid/gas and velocity vector, $P$ the pressure, $\mu$ the dynamic viscosity. And $h_i$ is the enthalpy of $i^{th}$ species, $T$ the temperature, $\lambda$ the thermal conductivity, $c_p$ the heat capacity at constant pressure, $\dot{\omega}_i$ the mole-based net production rate of $i^{th}$ species, $M_i$ the molecular weight of species $i$, and $\boldsymbol{J}_i$ the diffusion flux of $i^{th}$ species, $Y_i$ the mass fraction of $i^{th}$ species. The diffusion flux $\boldsymbol{J}_i$ is modeled based on Fick's law and mix-averaged correction:

$$\boldsymbol{J}_i = -\rho D'_i \nabla Y_i + \rho Y_i \sum_k D'_k \nabla Y_k \tag{5}$$

The $D'$ is the diffusion coefficients.

The chemical production rate is calculated from elementary reactions:

$$\sum_i^{N_S} v'_{ij} X_i \underset{K_{r,i}}{\overset{K_{f,j}}{\rightleftharpoons}} \sum_i^{N_S} v''_{ij} X_i, \quad X_i = \frac{Y_i}{M_i} \tag{6}$$

Where $v'_{ij}$ and $v''_{ij}$ are the stoichiometric coefficients of species $i$ in reaction $j$, respectively, $X_i$ the mole fraction of species $i$. $K_{f,j}$ and $K_{r,j}$ are the forward and reverse rate constant of species $i$ in reaction $j$, which are determined by Arrhenius equations:

$$K_{f,j} = A_j T^{\beta_j} \exp(-\frac{Ea_j}{RT}) \tag{7}$$

$$K_{r,j} = \frac{A_j T^{\beta_j} \exp\left(-\frac{Ea_j}{RT}\right)}{\exp\left(-\frac{\Delta G_j^0}{RT}\right)(RT)^{-\Sigma_i(v_{ij}'' - v_{ij}')}} \tag{8}$$

Here, $A$ is the pre-exponential factor (also called Arrhenius constant), $\beta$ the temperature exponent, $Ea$ the activation energy, $R$ the universal gas constant, $\Delta G^0$ the standard Gibbs free energy change. Thus, the net production rate $\dot{\omega}_i$ is:

$$\dot{\omega}_i = M_i \sum_{j=1}^{N_R} \left[(v_{ij}'' - v_{ij}')\left(K_{f,j}\prod_{i=1}^{N_S} C_i^{v_{ij}'} - K_{r,j}\prod_{i=1}^{N_S} C_i^{v_{ij}''}\right)\right], \quad C_i = \frac{\rho Y_i}{M_i} \tag{9}$$

Among the governing equations, chemical effect mainly works on the species and energy equations, and strongly affect the results in species distribution. With mass conservation in Eq. (1), the species conservation equations can be written as:

$$\rho \frac{\partial Y_i}{\partial t} + \underbrace{\rho \boldsymbol{U} \cdot \nabla Y_i}_{\text{Convection}} = \underbrace{\mathcal{S}_{\text{diff}}^Y(Y_i)}_{\text{Diffusion}} + \underbrace{M_i \dot{\omega}_i}_{\text{Reaction}} \tag{10}$$

Where the $\mathcal{S}_{\text{diff}}^Y(Y_i) = -\nabla \cdot \boldsymbol{J}_i$ represent the source term from diffusion in the right hand side of Eq. (3), respectively. Now that this formular stains with nonlinear convection term $\rho \boldsymbol{U} \cdot \nabla Y_i$, a transfer from R-D-C system to R-D system can be achieved by switch Eulerian frame to Lagrangian frame based on the residence time $\tau$ along with the streamline whose arc is $\ell$:

$$\boldsymbol{U} \cdot \nabla Y_i = |\boldsymbol{U}|\frac{dY_i}{d\ell} \tag{11}$$

$$d\tau = \frac{d\ell}{|\boldsymbol{U}|} \Rightarrow \boldsymbol{U} \cdot \nabla Y_i = \frac{dY_i}{d\tau} \tag{12}$$

Following the same strategy, the species and energy transport equation of a R-D-C system can be transferred to a R-D system as:

$$\rho \frac{dY_i}{d\tau} = \mathcal{S}_{\text{diff}}^Y(Y_i) + M_i \dot{\omega}_i + \mathcal{R}_{\text{time}}^Y(Y_i) \tag{13}$$

$$\rho c_p \frac{dT}{d\tau} = \mathcal{S}_{\text{diff}}^T(T) - \sum_{i=1}^{N_S} M_i h_i \dot{\omega}_i + \mathcal{R}_{\text{time}}^T(T) \tag{14}$$

Where the $\mathcal{S}_{\text{diff}}^T(T) = \nabla \cdot (\lambda \nabla T) - \nabla \cdot \left(\sum_{i=1}^{N_S} h_i \boldsymbol{J}_i\right)$ is the diffusion source term in the right hand side of energy conservation Eq. (4). $\mathcal{R}_{\text{time}}^Y(Y_i) = -\rho \frac{dY_i}{dt}$ and $\mathcal{R}_{\text{time}}^T(T) = -\rho c_p \frac{dT}{dt}$ represents the time variant residual of species and temperature in Eulerian frame, respectively.

In most cases in chemical studies, the experiments are often collected from a steady or quasi-steady state for stable solution, which means the time variant in Eulerian frame approaches zero: $\frac{d(\cdot)}{dt} \to 0, \mathcal{R}_{\text{time}} \to 0$. So, the R-D system can be solved by the following equations in the Lagrangian context as:

$$\rho \frac{dY_i}{d\tau} = \underbrace{\mathcal{S}_{\text{diff,sol}}^Y(Y_i)}_{\text{Diffusion}} + \underbrace{M_i \dot{\omega}_i}_{\text{Reaction}} \tag{15}$$

$$\rho c_p \frac{dT}{d\tau} = \underbrace{\mathcal{S}_{\text{diff,sol}}^T(T)}_{\text{Diffusion}} + \underbrace{(-\sum_{i=1}^{N_S} M_i h_i \dot{\omega}_i)}_{\text{Reaction}} \tag{16}$$

In a traditional and widely applied way, the ideal Lagrangian reactor in a pure chemical ODE raised assumptions of perfect stirred and uniform mixed in the Lagrangian particle. This would lead to $\nabla(\cdot) \to 0$, $\mathcal{S}_{\text{diff}} \to 0$, and a pure reaction ODE system as:

$$\rho \frac{dY_i}{d\tau} = \underbrace{M_i \dot{\omega}_i}_{\text{Reaction}} \tag{17}$$

$$\rho c_p \frac{dT}{d\tau} = \underbrace{-\sum_{i=1}^{N_S} M_i h_i \dot{\omega}_i}_{\text{Reaction}} \tag{18}$$

### 2.2. Diffusion-chemistry coupled Neural ODEs

Following the instinct from pure reaction ODE system, it's natural to couple neural ordinary differential equations into the iteration, and make the neuron represents every time differential part of the variable. Take example of species mass fraction, which is directly calculated from chemical reaction, the universal neuron $\mathcal{N}$ at time step $n$ and the iteration of species will be:

$$\frac{dY_i}{d\tau}(Y_i^n, \tau^n) = \mathcal{N}(Y_i^n, \mathcal{T}^n) \tag{19}$$

$$Y_i^{n+1} = \text{ODEsolver}\big(\mathcal{N}(Y_i^n), Y_i^0\big) \tag{20}$$

However, this kind of universal neuron structure can not capture the chemistry directly through mechanism since it only contains single state variable about species. From Arrhenius law, we know that the time variants are strongly depended by mechanism and Arrhenius coefficients $\boldsymbol{\theta}$ (represents the integrated contribution of $A$, $\beta$, Ea), so it's hard to get accessible and continuous information through such universal neuron. Previous CRNN method [37] directly implements the Arrhenius law into the neuron for stable convergence: $\mathcal{N} = \mathcal{N}^{\text{Chem}}\big(Y_i^n, \tau^n; (T^n, \boldsymbol{\theta})\big)$. But this structure can only be applied in reaction ODE. Thus, we proposed the diffusion-chemistry coupled structure of Diff-Chem neuron as:

$$\frac{dY_i}{d\tau}(Y_i^n, \tau^n) = \mathcal{N} = \mathcal{N}^{\text{Diff-Chem}}\big(Y_i^n, \tau^n; (T^n, P^n, \mathcal{S}_{\text{diff}}^n, \boldsymbol{\theta})\big) \tag{21}$$

Fig. 1 shows the integrated structure of Diff-Chem Neural ODE, where the single Arrhenius layer provides the reaction rate for one of the reactions, and the reaction layer gives time-differentiated outcomes (presented by $\boldsymbol{\phi}$ here) coupled with diffusion terms. The integration of this ODE system is along the residence direction from stream inlet (corresponding to $\tau^0$) to stream outlet (corresponding to $\tau^M$). Note that the forced state variable temperature and pressure here is for defining reaction rates and densities, and these will not affect the derivative process from time-differentiated outcomes to the chemical mechanism since the Arrhenius-reaction layers coupled the mechanism's contribution directly. This structure has several characteristics as:

- **End-to-End differentiation for chemical mechanism**: the Arrhenius-reaction layers encode mechanistic contributions explicitly, enabling direct and continuous gradients with respect to kinetic coefficients and reaction pathways, with reaction-level interpretability.
- **Stiffness-aware numerical integration**: the Arrhenius form parameterization constrains the sensitivity of chemical terms, while the flux-forced diffusion term avoids entangling transport effects with chemistry, improving numerical stability and allowing robust residence-direction integration with stiff-capable ODE solvers.
- **Pretraining-free**: the mechanism-embedded inductive bias and structured parameterization directly as the learnable weights of neurons, allowing training from scratch without staged pretraining.
- **Streamline-interpreted PDE applicability**: by integrating along the residence direction and introducing diffusion as a forced flux term, Diff-Chem Neural ODE naturally extends Arrhenius-embedded chemical neural networks from reaction ODEs to R-D-C PDE surrogates.

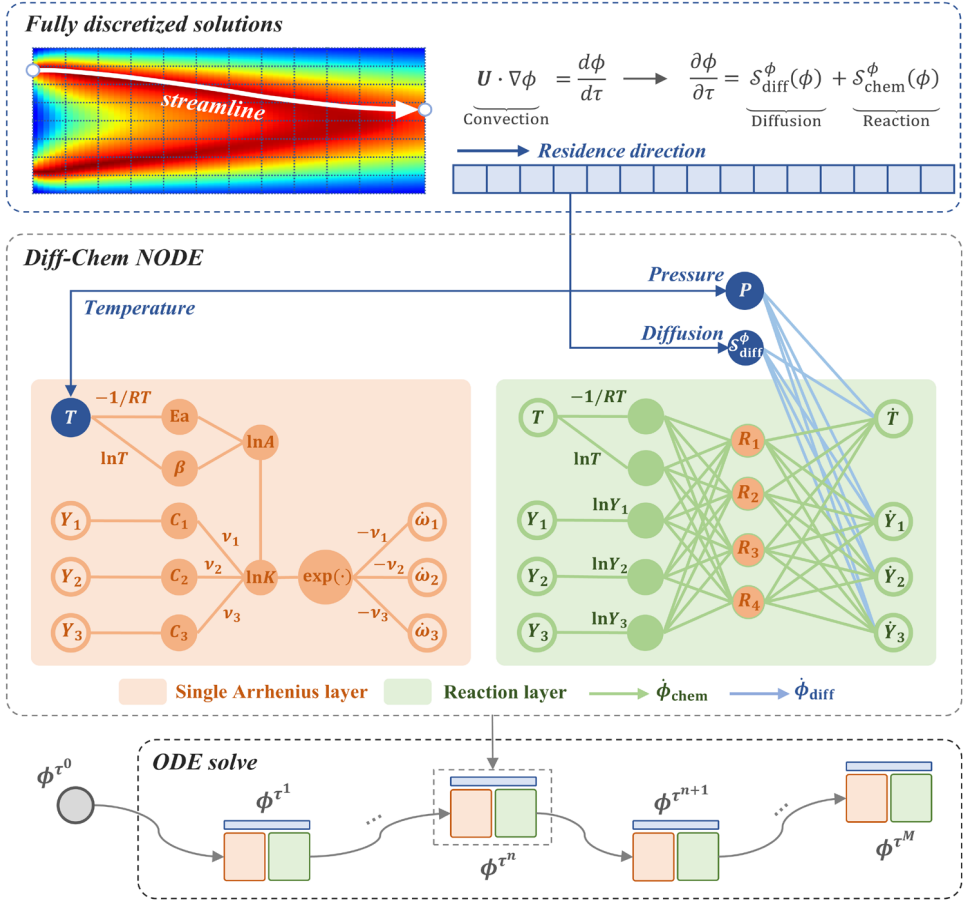

**Fig. 1. Integrated structure of Diff-Chem Neural ODE.**

*2.3. Differentiation and optimization strategy*

Let $\tau \in [\tau^0, \tau^M]$ denote the residence time coordinate, and the transported thermo-chemical state be $\boldsymbol{\phi}$ which contains $T$ and $\{Y_i\}_{i=1}^{N_S}$. The Diff-Chem ODE follows integration with solver operator $\psi$ as:

$$\frac{d\boldsymbol{\phi}}{d\tau} = \boldsymbol{f}(\boldsymbol{\phi}(\tau); \boldsymbol{\theta}) = \boldsymbol{S}_{\text{chem}}^{\phi}(\boldsymbol{\phi}(\tau); \boldsymbol{\theta}) + \boldsymbol{S}_{\text{diff}}^{\phi}(\tau) \tag{22}$$

$$\boldsymbol{\phi}(\tau^M) = \psi(\boldsymbol{\phi}(\tau^0), \boldsymbol{f}(\cdot; \boldsymbol{\theta}), \tau^0, \tau^M) \tag{23}$$

Consider the objective function to be the minimum loss $L$ with reference fields $\widehat{\boldsymbol{\phi}}$ sampled at $\{\tau^m\}_{m=0}^{M}$:

$$L(\boldsymbol{\theta}) = e\left(\boldsymbol{\phi}(\tau^m), \widehat{\boldsymbol{\phi}}(\tau^m)\right) \tag{24}$$

Where $e(\cdot)$ represents the mismatch such as Mean Absolute Error (MAE) or Mean Squared Error.

We employ two complementary differentiation routes: forward mode and adjoint/reverse mode. Forward mode AD propagates tangents from inlet to outlet along the residence marching direction. Adjoint mode using an interpolating adjoint strategy, and the key object is the adjoint propagated from outlet to inlet [48].

Define the first order gradient matrix $\boldsymbol{G} = \frac{\partial \boldsymbol{\phi}}{\partial \boldsymbol{\theta}}$. Differentiating Diff-Chem ODE yields standard forward gradients system:

$$\frac{d}{d\tau}\left(\frac{\partial \boldsymbol{\phi}}{\partial \boldsymbol{\theta}}\right) = \frac{d}{d\boldsymbol{\theta}}\left(\frac{d\boldsymbol{\phi}}{d\tau}\right) = \frac{d\boldsymbol{f}}{d\boldsymbol{\theta}} = \frac{\partial \boldsymbol{f}}{\partial \boldsymbol{\phi}}\frac{\partial \boldsymbol{\phi}}{\partial \boldsymbol{\theta}} + \frac{\partial \boldsymbol{f}}{\partial \boldsymbol{\theta}} \tag{25}$$

$$\frac{d\boldsymbol{G}}{dt} = \underbrace{\frac{\partial \boldsymbol{f}}{\partial \boldsymbol{\phi}}}_{\boldsymbol{\mathcal{J}}_\phi(\tau)} \boldsymbol{G} + \underbrace{\frac{\partial \boldsymbol{f}}{\partial \boldsymbol{\theta}}}_{\boldsymbol{\mathcal{J}}_\theta(\tau)} \tag{26}$$

Where the Jacobian matrix $\boldsymbol{\mathcal{J}}_\phi$ and $\boldsymbol{\mathcal{J}}_\theta$ can be directly obtained through AD technique. Thus, the forward mode just needed for integrating an augmented ODE

$$\frac{d}{d\tau}\begin{bmatrix} \boldsymbol{\phi} \\ \text{vec}(\boldsymbol{G}) \end{bmatrix} = \begin{bmatrix} \boldsymbol{f}(\boldsymbol{\phi}(\tau); \boldsymbol{\theta}) \\ \text{vec}\left(\boldsymbol{\mathcal{J}}_\phi(\tau)\boldsymbol{G} + \boldsymbol{\mathcal{J}}_\theta(\tau)\right) \end{bmatrix} \tag{27}$$

with stiff-capable time stepping consistent with the primal solve. This is why forward sensitivities remain stable for stiff source terms. So, the gradients for objective function under forward mode is:

$$\frac{\partial L}{\partial \boldsymbol{\theta}} = \frac{\partial e}{\partial \boldsymbol{\phi}} \boldsymbol{G} \tag{28}$$

Although the forward mode is clean and stable, when parameter dimension is large, the forward mode will become expensive since it's number of ODE scales linearly in the number of parameters times state variables [38]. So, the adjoint/reverse mode is also employed here for increasing efficiency in large number of parameters. And for a stable convergence in both non-stiff and stiff case, we used interpolating adjoint method for stable convergence [36].

Define an adjoint variable **a**

$$\mathbf{a}(\tau) = \frac{\partial L}{\partial \boldsymbol{\phi}(\tau)} \tag{29}$$

The adjoint evolves backward between observation points according to:

$$\frac{d\mathbf{a}}{\tau} = -\boldsymbol{\mathcal{J}}_\phi(\tau)^\top \mathbf{a}(\tau) \tag{30}$$

$$\mathbf{a}(\tau^{m-}) = \mathbf{a}(\tau^{m+}) + \frac{\partial e}{\partial \boldsymbol{\phi}}\bigg|_{\phi(\tau^m)} \tag{31}$$

Thus, the gradient is then given by:

$$\frac{\partial L}{\partial \boldsymbol{\theta}} = \int_{\tau^0}^{\tau^M} \mathbf{a}(\tau)^\top \boldsymbol{\mathcal{J}}_\theta(\tau)\, d\tau \tag{32}$$

Define the integration state vector function **g**:

$$\mathbf{g}(\tau) = \int_{\tau}^{\tau^M} \mathbf{a}(\tau)^\top \boldsymbol{\mathcal{J}}_\theta(\tau) \, d\tau \tag{33}$$

$$\frac{d\mathbf{g}}{\tau} = -\mathbf{a}(\tau)^\top \boldsymbol{\mathcal{J}}_\theta(\tau) = -\boldsymbol{\mathcal{J}}_\theta(\tau)^\top \mathbf{a}(\tau) \tag{34}$$

So that $\mathbf{g}(\tau^M) = 0$ and $\mathbf{g}(\tau^0) = \int_{\tau^0}^{\tau^M} \mathbf{a}(\tau)^\top \boldsymbol{\mathcal{J}}_\theta(\tau) \, d\tau$, and the gradient becomes:

$$\frac{\partial L}{\partial \boldsymbol{\theta}} = \mathbf{g}(\tau^0) \tag{35}$$

Thus, the adjoint mode is achieved by solving adjoint augmented system as:

$$\frac{d}{d\tau}\begin{bmatrix}\mathbf{a}\\ \mathbf{g}\end{bmatrix} = \begin{bmatrix}-\boldsymbol{\mathcal{J}}_\phi(\tau)^\top \mathbf{a}(\tau)\\ -\boldsymbol{\mathcal{J}}_\theta(\tau)^\top \mathbf{a}(\tau)\end{bmatrix}, \ \mathbf{a}(\tau^{M+}) = 0, \ \mathbf{g}(\tau^M) = 0 \tag{36}$$

A practical challenge in stiff ODEs is that reverse integration needs access to $\boldsymbol{\phi}(\tau)$ when evaluating $\boldsymbol{\mathcal{J}}_\phi(\tau)$ and $\boldsymbol{\mathcal{J}}_\theta(\tau)$. The interpolating adjoint strategy addresses this by constructing a continuous approximation $\widetilde{\boldsymbol{\phi}}(\tau)$ from the forward solution:

1) Forward pass (primal solve): integrate Eq. (22) and (23) and store states at solver steps $\{(\tau^m, \boldsymbol{\phi}^m)\}_{m=0}^{M}$;
2) Dense interpolation by selected method $M$: $\widetilde{\boldsymbol{\phi}}(\tau) = M(\tau; \{(\tau^m, \boldsymbol{\phi}^m)\}_{m=0}^{M})$. where $M$ can be a piecewise polynomial (e.g., Hermite/cubic) consistent with the ODE solver's dense output.

So, the Jacobian matrix can be given by $\boldsymbol{\mathcal{J}}_\phi(\tau) = \left.\frac{\partial f}{\partial \boldsymbol{\phi}}\right|_{\boldsymbol{\phi}=\widetilde{\boldsymbol{\phi}}(\tau)}$ and $\boldsymbol{\mathcal{J}}_\theta(\tau) = \left.\frac{\partial f}{\partial \boldsymbol{\theta}}\right|_{\boldsymbol{\phi}=\widetilde{\boldsymbol{\phi}}(\tau)}$.

After differentiation, the gradients of objective function are obtained, and the optimization could be directly implemented through gradient-based optimization. Denote the gradient-based optimizer as $\eta$, the update of parameters can be written as:

$$\boldsymbol{\theta}^{k+1} = \boldsymbol{\theta}^k + \eta\left(\boldsymbol{\theta}^k, \left.\frac{\partial L}{\partial \boldsymbol{\theta}}\right|_{\boldsymbol{\theta}^k}\right) \tag{37}$$

The optimizer is the adaptive moment estimation (Adam) [49] for its fast and stable convergence in stochastic gradient descent tasks, and the learning rate is set as 0.01.

**Algorithm 1: Pseudocode for Diff-Chem Neural ODE based optimization strategy**

| Block | | Pseudocode | Key outputs |
|---|---|---|---|
| 1 | **Inputs** | Observation data $\widehat{\boldsymbol{\phi}}$ as reference; Initial chemical states $\boldsymbol{\theta}^0$; Optimizer setting $\eta$ (Adam); Maximum iteration steps $N_{\text{opt}}$; | - |
| 2 | **Initialization of forced profiles** | Compute forced profiles $T$, $P$, $\boldsymbol{S}_{\text{diff}}^\phi$ from initial fully discretized solutions | $T$, $P$, $\boldsymbol{S}_{\text{diff}}^\phi$ |
| 3 | **Diff-Chem model** | Define RHS: $\boldsymbol{f}(\boldsymbol{\phi}(\tau);\boldsymbol{\theta}) = \boldsymbol{S}_{chem}^\phi(\boldsymbol{\phi}(\tau);\boldsymbol{\theta}) + \boldsymbol{S}_{diff}^\phi(\tau)$; Define loss: $L(\boldsymbol{\theta}) = e\left(\boldsymbol{\phi}(\tau^m), \widehat{\boldsymbol{\phi}}(\tau^m)\right)$ | $\boldsymbol{f}$, $L^0$ |
| 4 | **Iteration** | For $k=0,1, \ldots, N_{\text{opt}}-1$, **do** | |
| | 4.1 Primal solve | Solve $\{\tau^m, \boldsymbol{\phi}^m\}_m^M \leftarrow \boldsymbol{\phi}(\tau^M) = \psi(\boldsymbol{\phi}(\tau^0), \boldsymbol{f}, \tau^0, \tau^M)$ by ODE solver $\psi$ (e.g., BDF/TRBDF [50]) | $\{\tau^m, \boldsymbol{\phi}^m\}_m^M$ |
| | 4.2 Evaluate loss | $\boldsymbol{\phi}^m \leftarrow$ (i) forward mode: direct readout; else (ii) adjoint mode: interpolate $\widetilde{\boldsymbol{\phi}}(\tau)$; Compute loss $L^k$ | $L^k$ |
| | 4.3 Gradients | $\nabla_{\boldsymbol{\theta}^k} L^k \leftarrow \boldsymbol{\mathcal{J}}_\phi^k, \boldsymbol{\mathcal{J}}_\theta^k$: Propagate gradients by selected AD methods; | $\nabla_{\boldsymbol{\theta}^k} L^k$ |

| | | Integrate augment ODEs | |
|---|---|---|---|
| 4.4 | Update | $\boldsymbol{\theta}^{k+1} \leftarrow \boldsymbol{\theta}^k + \eta(\boldsymbol{\theta}^k, \nabla_{\boldsymbol{\theta}^k} L^k)$ , update kinetic parameters; $[T^{k+1}, P^{k+1}, \boldsymbol{S}_{\text{diff}}^{\phi,k+1}] \leftarrow \boldsymbol{\theta}^{k+1}$, update profiles | $\boldsymbol{\theta}^{k+1}, T^{k+1},$ $P^{k+1}, \boldsymbol{S}_{\text{diff}}^{k+1}$ |
| 4.5 | Stopping | **if** convergence (e.g., $\|\nabla L\|$small, or max iters) **then break**. | |
| 5 | Return | - | Final $\boldsymbol{\theta}^{\text{opt}}$ |

As conducted from above, the overall procedure can be summarized as an iterative "simulate-compare-update" loop built around a differentiable reaction-diffusion system, which has been shown in Algorithm 1. First, it takes measured/high-fidelity) trajectories as the reference and initializes the unknown kinetic/chemical parameters with a reasonable initial guess. Any externally prescribed conditions that drive the dynamics are prepared so that the model can be evaluated consistently over time. Given these inputs, the system evolution is then generated by integrating the ODE forward in time with a solver suitable for both non-stiff and stiff chemistry, producing a predicted state trajectory. This prediction is compared against the reference data at the available observation times, and the discrepancy is aggregated into a scalar objective function. Because the simulator is differentiable end-to-end, gradients of the objective with respect to the kinetic parameters can be obtained. Finally, an optimizer uses these gradients to update the parameters, and the cycle repeats until the loss stops improving or a preset iteration budget is reached. The output is a calibrated parameter set that best aligns the simulated dynamics with the observed behavior.

We implemented the whole workflow into two versions based on open-source platforms in Just-In-Time (JIT) language Julia and Python language. In Julia language, we mainly used open-source platform *Arrhenius.jl* [51] and *DifferentialEquations.jl* [52] for chemical calculation and differential equations. While in Python language, we mainly used a C++ complied open-source platform *Cantera* [53] with mature integrated differentiable chemical framework, and *Scikit-Learn* [54], *Pytorch* [55] platforms for gradients computation and mature optimizers. The ODE solver is coupled in the platforms, and we mainly use the stiff-robust solvers [50]. Since the computation in Python-based version can be more stable for *Cantera*'s high robustness in species normalization, we use Julia-based version for only low stiffness cases while Python-based version and forward mode AD for higher stiffness cases.

## 3. Demonstration in reacting flows

In this study, two canonical burner-stabilized reacting-flow configurations are selected as representative gas-phase oxidation cases for validation, with a focus on mechanism optimization. Oxidation under combustion-relevant conditions features intense heat release and pronounced kinetic stiffness (wide time-scale separation), which provides a stringent and practically relevant benchmark for verifying the robustness and effectiveness of the proposed methods. Fig. 2 shows the configurations.

The Burner-Stabilized Flat (BSF) reacting flow provides a chemically well-defined environment with minimal geometric complexity, making it suitable for assessing detailed reaction mechanisms and transport models under controlled residence time and thermal conditions. The distributed stabilization by a porous/plenum-fed burner yields an approximately planar reaction zone over the central region, and the dominant gradients are aligned with the burner-normal direction, so the configuration is often treated using a quasi-one-dimensional description for the core region. This case is selected because it isolates chemistry and molecular transport effects (e.g., differential diffusion and heat release) while avoiding uncertainties associated with multidimensional flame curvature and complex aerodynamics. Boundary conditions are imposed at the burner exit (inlet) by prescribing the mixture composition (species mass fractions or mole fractions), inlet temperature, and either mass flux or axial velocity (equivalently, the inlet Reynolds number), together with the operating pressure. The burner surface can be modeled as isothermal at a specified burner temperature or as a prescribed heat-loss boundary (where burner heat conduction is represented). At the downstream boundary, a pressure outlet is typically applied with convective (zero-gradient) conditions for species and temperature, ensuring a non-reflecting outflow as the solution approaches the fully reacted/post-reacting state.

The Burner-Stabilized Stagnation (BSS) reacting flow is selected as a complementary canonical case because it introduces a controlled strain field and strong flow-chemistry interaction while retaining a clean, well-characterized geometry. An axisymmetric jet issued from the burner decelerates toward a stagnation surface, producing a well-defined stagnation point and an adjustable strain rate, which enables systematic interrogation of kinetics-transport coupling and, when relevant, extinction/near-extinction behavior. Near the centerline the structure is commonly represented along the stagnation streamline, which motivates reduced-order (quasi-1D) treatments without changing the physical interpretation of the boundary conditions. Boundary conditions include prescribed inlet composition, inlet temperature, and inlet velocity (or mass flux) at the burner exit, together with the operating pressure; symmetry on the axis ($U_r$=0 and zero radial gradients of scalar quantities); and wall conditions at the stagnation plate such as no-slip for velocity and a specified thermal condition (isothermal wall temperature or adiabatic wall), with an impermeable boundary for species (zero normal diffusive flux) unless heterogeneous chemistry or wall transpiration is considered. The outer/far-field boundary is typically treated as a pressure outlet (or specified co-flow when present) with convective/zero-gradient conditions for transported scalars to avoid artificially constraining the reacting flow.

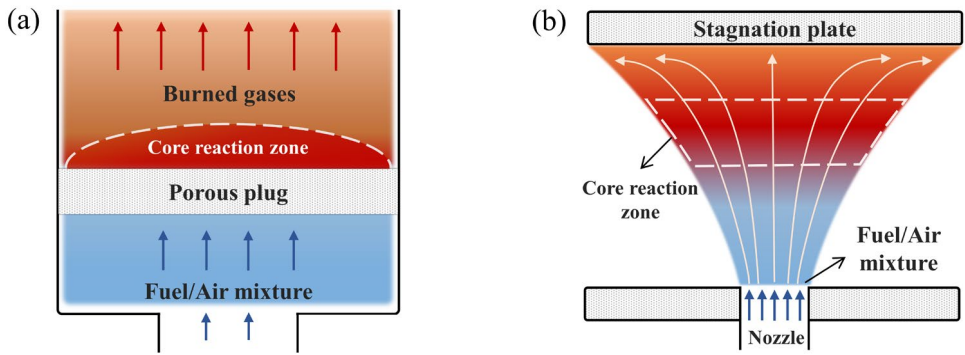

**Fig. 2. Configurations of (a) Burner-stabilized flat, and (b) Burner-stabilized stagnation flames.**

For comprehensive study of current method in different scenario, three representative mechanisms are selected here for validations, as shown in Table 1. The selected set spans a wide range of chemical complexity and numerical stiffness, from a small hydrogen combustion mechanism (Li-H2, 9 species/19 reactions; low stiffness) to two methane/hydrocarbon combustion mechanisms of increasing size (DRM19, 21 species/84 reactions; moderate stiffness; and GRI3.0, 53 species/325 reactions; high stiffness). All mechanisms are evaluated under the same two burner-stabilized reacting-flow configurations (BSF and BSS), so that the influence of mechanism size and stiffness on the optimization performance can be assessed consistently across both unstrained and strained oxidation environments.

Table 1: Mechanism summary

| Mechanism name | Species | Reactions | Fuel type | Test flow | Stiffness level |
|---|---|---|---|---|---|
| Li-H2 | 9 | 19 | Hydrogen | BSF&BSS | Low |
| DRM19 | 21 | 84 | Methane (Hydrocarbon) | BSF&BSS | Moderate |
| GRI3.0 | 53 | 325 | Methane (Hydrocarbon) | BSF&BSS | High |

*3.1. Comparison in accuracy*

As mentioned, the Diff-Chem model can directly predict the whole R-D system, while the previous pure chemical model which based on perfect mixed assumption can not give a comprehensive representation. So, we first conduct verifications of the Diff-Chem model based on cases listed in Table 1, and the results of BSF flame and BSS flame are shown in Fig. 3 and Fig. 4, respectively.

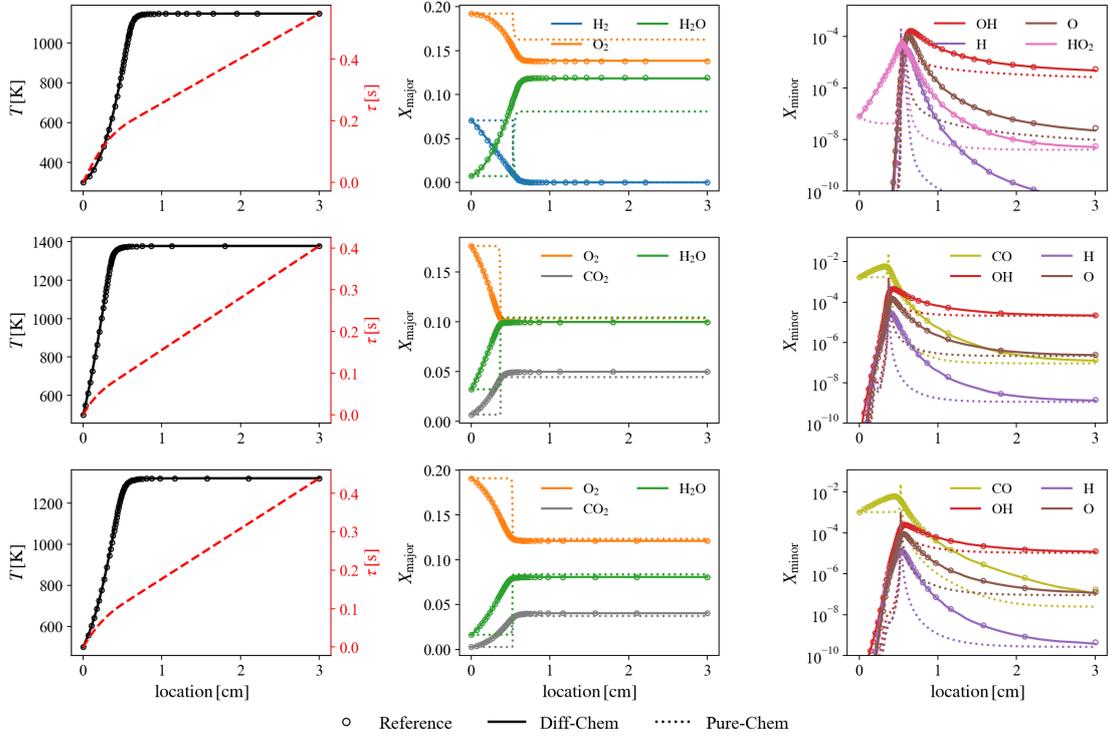

**Fig. 3. Comparison of different methods for BSF flames using Li-H2 mechanism (first row), DRM19 mechanism (second row), and GRI3.0 mechanism (last row).**

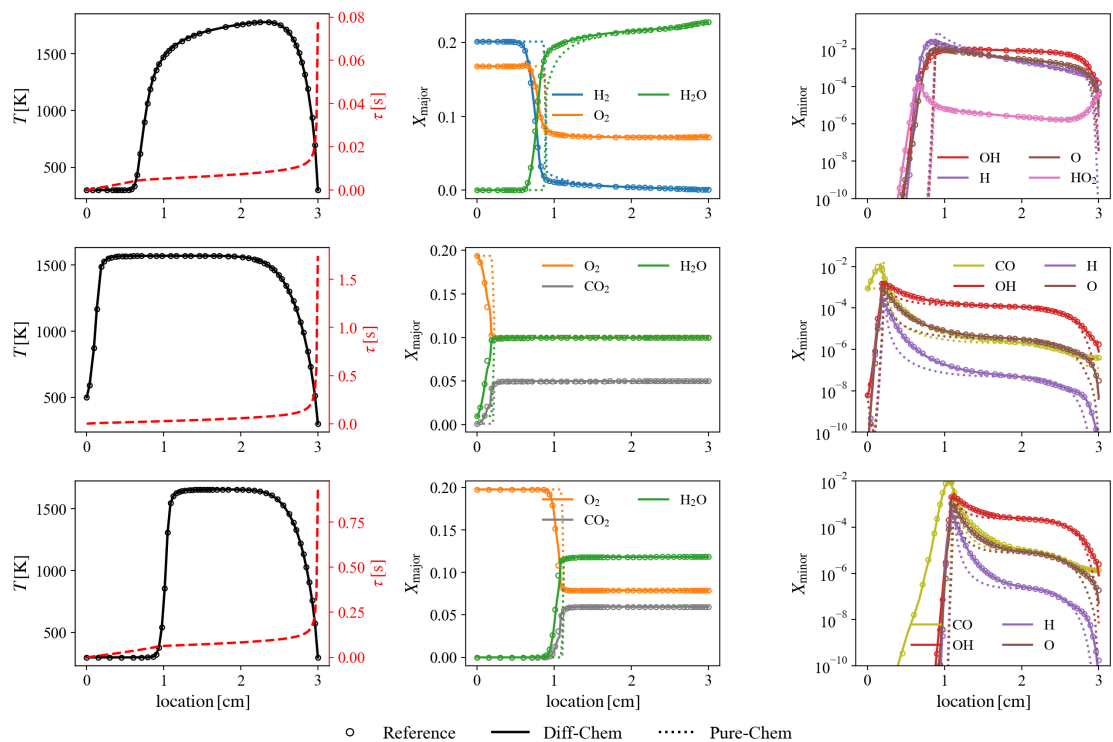

**Fig. 4.** Comparison of different methods for BSS flames using Li-H2 mechanism (first row), DRM19 mechanism (second row), and GRI3.0 mechanism (last row).

In these two figures, the first column presents the base temperature and residence-time profiles for the current cases, while the second and third columns report the predicted concentrations of representative major and minor species, respectively. First, the temperature and residence-time profiles show that the residence time in the BSS flame exhibits substantially stronger nonlinearity than that in the BSF flame due to the stagnation effect. This enhanced nonlinearity is accompanied by a reduction in temperature and species concentrations in the BSS configuration. Notably, the temperature predictions are identical for Diff-Chem Neural ODE and the pure-chemistry Neural ODE (hereafter, Pure-Chem Neural ODE), because both models use exactly the same state variables extracted from the full PDE solutions to ensure a fair comparison; consequently, the only distinction between the two models is whether diffusion terms are included (as illustrated in Fig. 1). In contrast, the species predictions reveal a pronounced discrepancy attributable to diffusion. Across all cases, Diff-Chem Neural ODE reproduces the full-PDE reference with essentially exact agreement, whereas Pure-Chem Neural ODE exhibits substantial deviations.

More importantly, Pure-Chem Neural ODE effectively neglects the preheat and reaction processes preceding intense combustion, thereby yielding an incorrect ignition location and an unrealistically thin reaction zone. This deficiency is critical for mechanism assessment: inaccurate gradients in an excessively thin flame region prevent reliable quantification of kinetic contributions both upstream of ignition and in the vicinity of the flame front. Upon closer inspection, for the low-stiffness hydrogen BSF flames based on the Li-H2 mechanism, Pure-Chem Neural ODE fails to predict even the post-flame major-species levels accurately, which in turn introduces large errors in reaction-rate calculations that depend directly on species composition. For the high-stiffness hydrocarbon flames governed by the DRM19 and GRI3.0 mechanisms, although post-flame errors are smaller than in the hydrogen cases, the minor species still exhibit an unphysically abrupt, near-discontinuous variation at the flame front.

For the BSS flame, the overall conclusions are consistent with those for the BSF flame, but an additional feature arises near the stagnation point. Pure-Chem Neural ODE tends to predict a premature "stagnation" behavior, because it is overly sensitive to local temperature decreases upstream of the stagnation point. In the diffusion-coupled system, however, the influence of temperature is moderated by transport and diffusive redistribution; incorporating diffusion therefore enables Diff-Chem Neural ODE to produce markedly improved predictions. We also observe that, at the stagnation point in the BSS flame, Diff-Chem Neural ODE slightly underpredicts minor-species levels relative to the reference. This discrepancy is attributable to the residence-time estimation at stagnation, where the velocity approaches zero and the conventional definition becomes singular; the residence time is instead inferred via continuity-based arguments and cannot be made perfectly exact. This issue does not affect the central conclusions, because i) minor-species concentrations in this region are already extremely small and thus not the primary focus, and ii) after reducing the BSS configuration to a centerline one-dimensional model, a known probe effect exists [56] —one typically shifts a few grid points upstream of the stagnation point to better represent the centerline of the original high-dimensional axisymmetric solution. In summary, Diff-Chem Neural ODE clearly outperforms Pure-Chem Neural ODE in directly predicting species composition: it faithfully captures diffusion-coupled effects and consistently reflects them in the resulting species profiles.

*3.2. Comparison in robustness and efficiency*

Above results show the accuracy of current diffusion-chemical coupled model compared with previous pure chemical model, while the robustness and efficiency of them can be compared further through disturbed initial states and gradients propagation tests. Fig. 5 gives an example of selected components' evolution along with ODE solver's iterative time steps. This compared point is the hydrogen BSF flame's last grid point, denoted as 'Target at $\tau^*$', and the x-axis indicate the ODE solver's internal time evolution when it is iterating for the objective time $\tau^*$. It can be seen that the components evolute more and more close to the final target by ODE solver, but the converged solution of pure chemical method shows large deviations from the target. The last column gives the evolution trajectory of components' absolute error, and it indicate that the Pure-Chem Neural ODE can not converge to the true value even after ODE solver's iteration. For Diff-Chem Neural ODE, the start error can be reduced and finally converge to the true target value.

Further, different levels of gaussian random disturbances $\mathcal{N}(0, \sigma^2)$ are added into the initial components' state to examine the performance under random initialization from fake states. Fig. 6 compares the error distributions for three noise levels, $\sigma$=0.05, 0.10 and 0.20, with 1,000 samples in each setting. The blue curves correspond to the diffusion-chemistry coupled method, whereas the orange curves represent the pure-chemistry method. From the statistical means, Diff-Chem Neural ODE consistently achieves smaller errors and thus higher accuracy. As the sampling bandwidth (noise level) increases, the error distributions of both methods broaden accordingly; however, the Pure-Chem Neural ODE exhibits a markedly larger and wider spread, indicating substantially higher sensitivity to initialization noise. This behavior is expected: perturbations in the initial state directly bias the reaction rates evaluated at the same nominal state, and such rate errors propagate through the ODE integration and accumulate over iterations. Overall, Diff-Chem Neural ODE maintains a uniformly smaller error across all noise levels.

Fig. 7 further reports the computational cost of several approaches for gradient evaluation. For a fair comparison, all tests are conducted under the same setting: the GRI3.0 mechanism (325 reactions) in a non-stiff hydrogen flame, and the reported quantity is the average gradient computed using forward finite differences. Under this protocol, every method must solve one baseline system and an additional 325 perturbed systems to obtain the gradient, ensuring an identical number of model evaluations across methods. As shown, the fully discretized PDE approach is the slowest, since each evaluation requires iterating the complete PDE system. Both Neural ODE-based approaches provide substantial speedups. Pure-Chem Neural ODE yields the lowest wall-clock time, because the purely chemical system is generally less stiff than the diffusion-coupled system and therefore requires fewer adaptive step-size reductions during ODE integration. However, when accuracy is taken into account, Diff-Chem Neural ODE offers the best overall trade-off, delivering both high computational efficiency and the most accurate results.

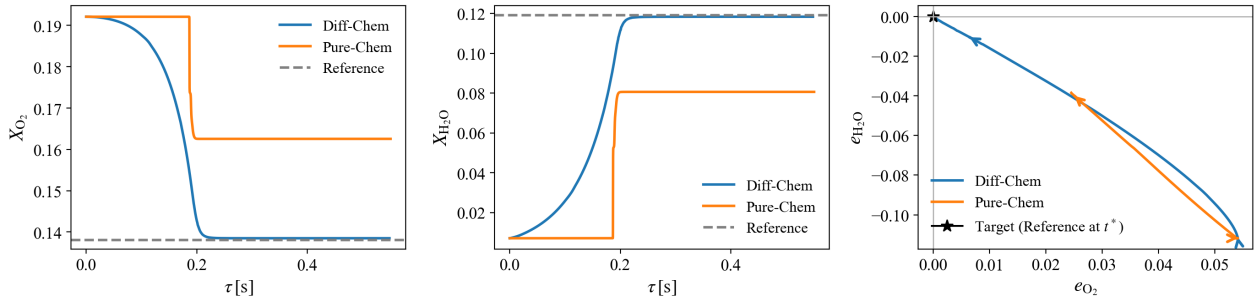

**Fig. 5. ODE internal evolution results of representative mole fractions (first and second column), and absolute error (third column).**

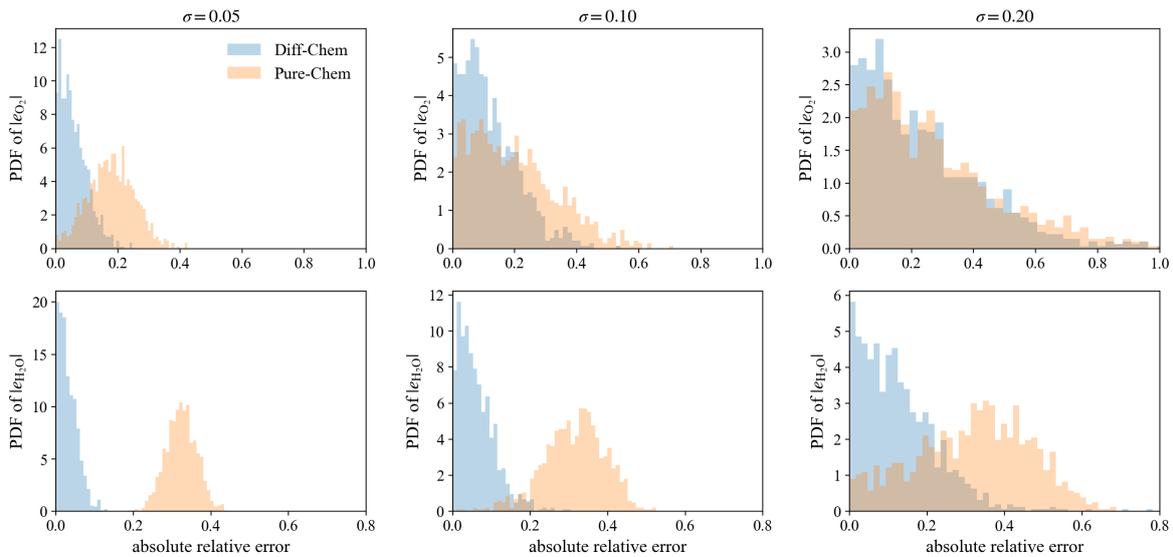

**Fig. 6. PDF of final absolute relative error under different disturbance levels: $\sigma$=0.05 (first column), $\sigma$=0.10 (second column), and $\sigma$=0.20 (third column).**

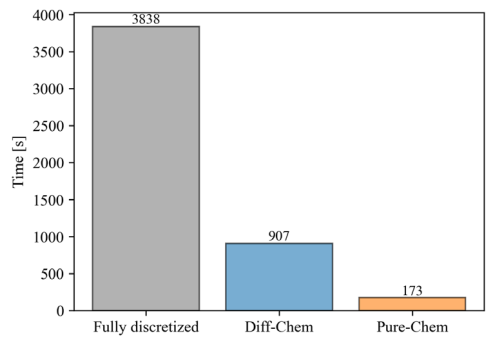

**Fig. 7. Gradient computational time of different methods.**

## 3.3. Kinetics optimization by Diff-Chem Neural ODEs

Diff-Chem Neural ODE leverages a differentiable architecture to directly obtain sensitivity contributions associated with chemical reactions, eliminating the need for additional sampling-based pre-training and enabling immediate use in gradient-based optimization. More importantly, this framework supports chemical path–based optimization under limited experimental observability (e.g., a restricted set of operating conditions and/or a limited number of measurable species), and therefore exhibits strong generalization capability. In what follows, we conduct optimization tests for the flames listed in Table 1.

For each flame, we uniformly sample 50 grid points along the streamwise direction and select several representative species as optimization targets, referred to here as primal species; the remaining species excluded from the objective are termed secondary species. We set up this 'only primal species'-oriented optimization because of that, in experiments, the detectable species are constrained by the current measurement techniques' limitations, which means the true values we can obtained for kinetic studies are limited to a few specific species. This setting can help verify the capabilities and effectiveness of current method under such scenarios. Besides, to account for the large differences in species magnitudes, we employ log-MSE as the optimization loss to balance scales across species. For $H_2$ flames, the primal species are $H_2$, $O_2$, and $H_2O$. For $CH_4$ flames, the primal species are $O_2$, $CO_2$, $H_2O$, $OH$, $CO$, $O$, and $H$. Following the Arrhenius law, we define the scaled reaction-rate parameters as the kinetic optimization inputs, $\theta = K/K_{ref}$, so that the ground-truth value is $\theta_{ref} = 1$. In the optimization tests, we initialize $\theta$ by randomly perturbing each component within [-0.5, 2], i.e., up to approximately twofold smaller or larger than the reference value.

Fig. 8 and Fig. 9 present the loss-evolution curves for the BSF and BSS flames, respectively. The horizontal axis denotes the updated state obtained after each optimization iteration, and we cap the optimization at 100 steps. For a more complete comparison, whenever the loss does not exhibit a clear increase, we allow the optimization to proceed to the maximum number of iterations. Across all cases, the loss decreases effectively and is ultimately reduced by more than 98%. Small oscillations are observed, which is expected: we do not tune the hyperparameters of gradient descent, but instead evaluate all cases under an identical set of hyperparameters. Nevertheless, the final loss values are close to convergence, with neither noticeable rebound nor renewed rapid decline, indicating that the optimization setup is stable and appropriate. More importantly, the losses of secondary species which are not concerned in optimized objects can also be reduced along the primal species. This shows the high generality of current Diff-Chem method for its effective optimization in unseen species.

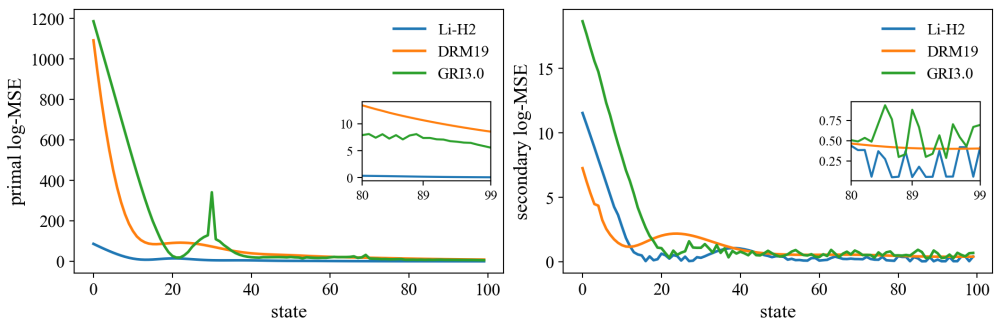

**Fig. 8. Losses evolution in BSF flames for primal species (left column) and secondary species (right column).**

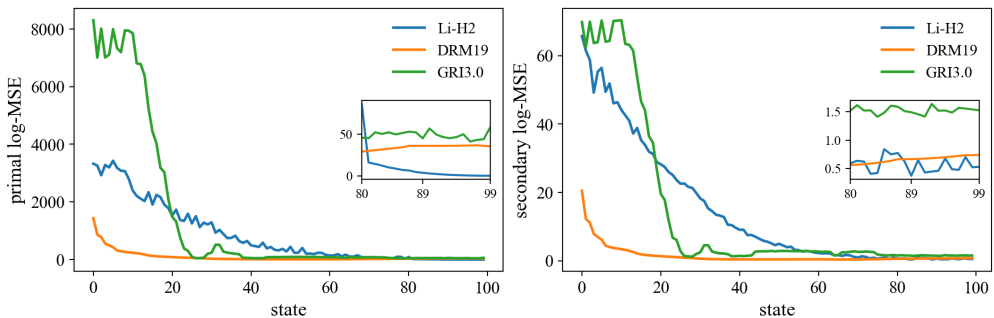

**Fig. 9. Losses evolution in BSS flames for primal species (left column) and secondary species (right column).**

Fig. 10 and Fig. 11 provide a detailed comparison of the solutions before and after optimization. For clarity, the plotted species mole-fraction profiles include only a subset of representative species selected from both the primal and secondary sets; the remaining species exhibit similar behavior. Moreover, we focus on regions with the most pronounced changes, as variations elsewhere are consistent but less visually distinctive. Overall, all variables evolve from the dashed initial state toward the final optimized state and ultimately match the target (reference) profiles almost exactly. Importantly, the optimization objective is defined only on the primal species, meaning that temperature and all secondary species are never directly "seen" by the gradient-based optimizer. Nonetheless, they also converge to the reference solution. This provides strong evidence that the physics-consistent neural formulation achieves effective global control of the coupled system. More importantly, the shift in the ignition location (core reaction zone) is also corrected accurately through optimization, further demonstrating the strong physical consistency and optimization capability of the proposed method.

In addition, to examine the consistency between gradient descent and the evolution of the kinetic inputs, Fig. 12 reports the trajectories of the kinetic-input scalars and the mean gradient magnitude for five representative reactions selected from the Li-H2 BSS flame. The results show that reactions associated with larger gradient magnitudes tend to be driven back toward the ground-truth value of 1, whereas those with small gradient magnitudes exhibit no clear monotonic trend and may fluctuate as they are indirectly affected by the updates of other, more influential reactions. This observation indicates that the directions of input updates are consistent with the gradient signals: larger gradients induce larger and more systematic corrections. These findings also suggest that, in the present system, reactions with small gradients have limited impact on the solution and can be regarded as low-sensitivity reactions. Such information can serve as a useful reference for subsequent kinetic refinement or targeted reaction optimization. At the same time, the observed behavior highlights the non-uniqueness of reaction-kinetic parameterization under specific structural constraints, i.e., multiple parameter combinations may produce comparably similar macroscopic flame responses.

In summary, these results collectively demonstrate the effectiveness and scalability of Diff-Chem Neural ODE for optimization tasks, enabling efficient, physically consistent optimization even under limited measurement conditions.

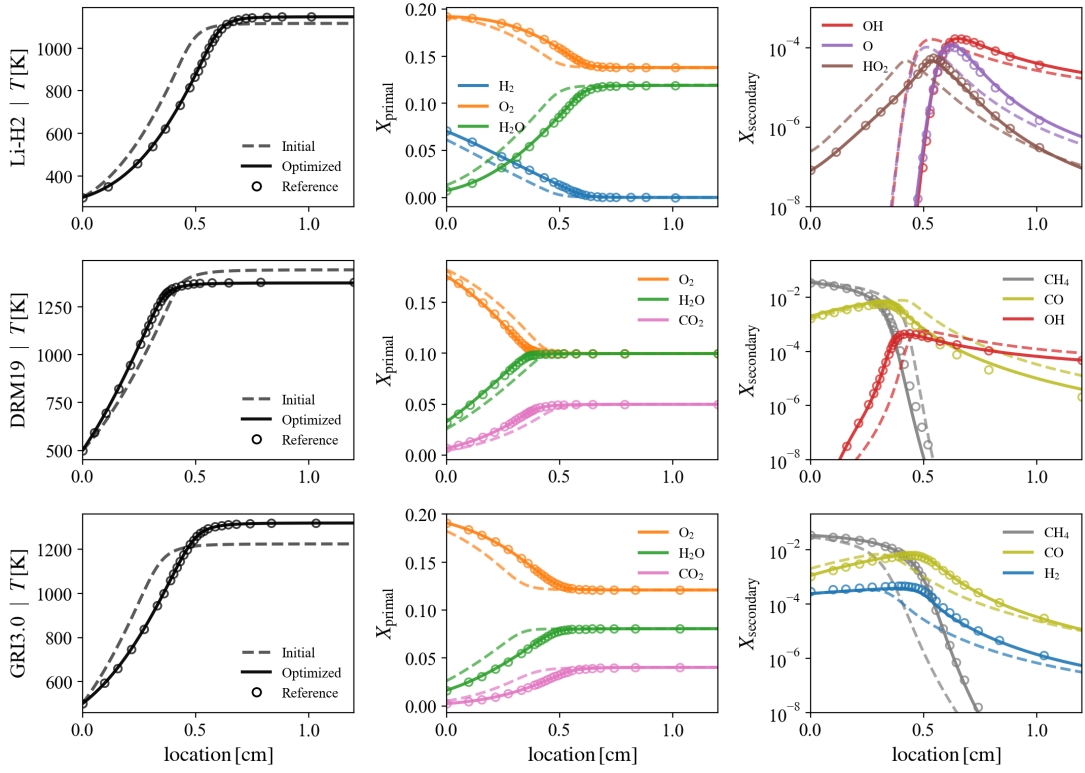

**Fig. 10.** Comparisons of initial and optimized states in BSF flames.

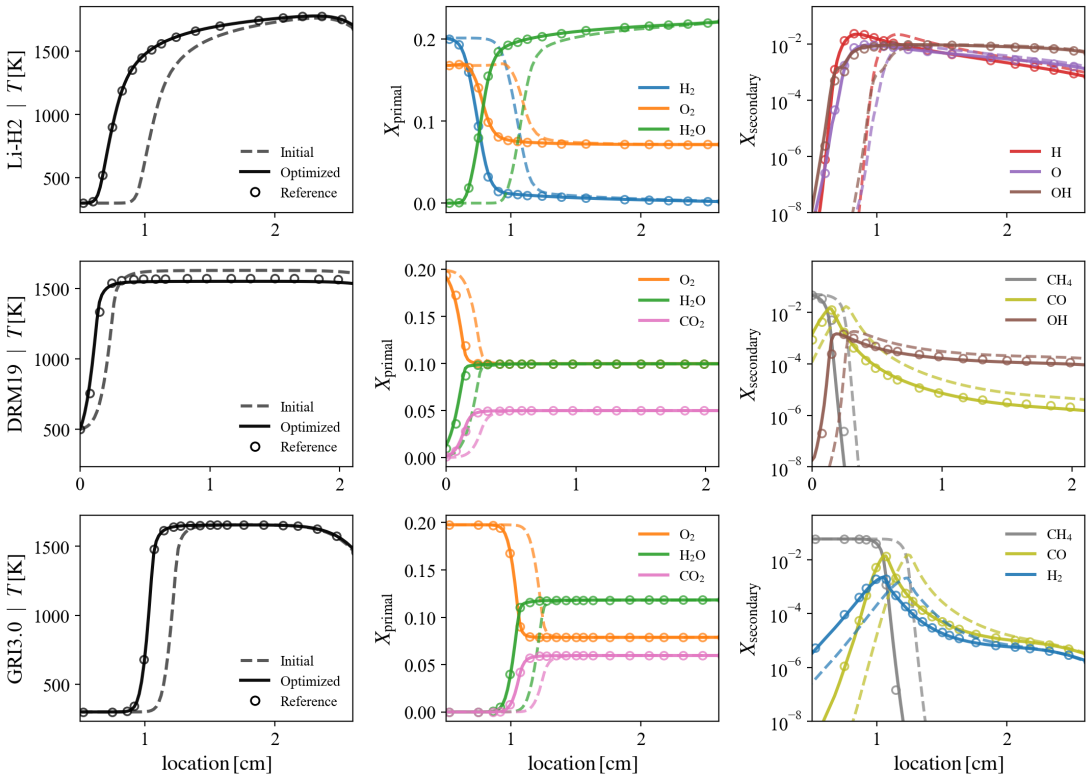

**Fig. 11.** Comparisons of initial and optimized states in BSS flames.

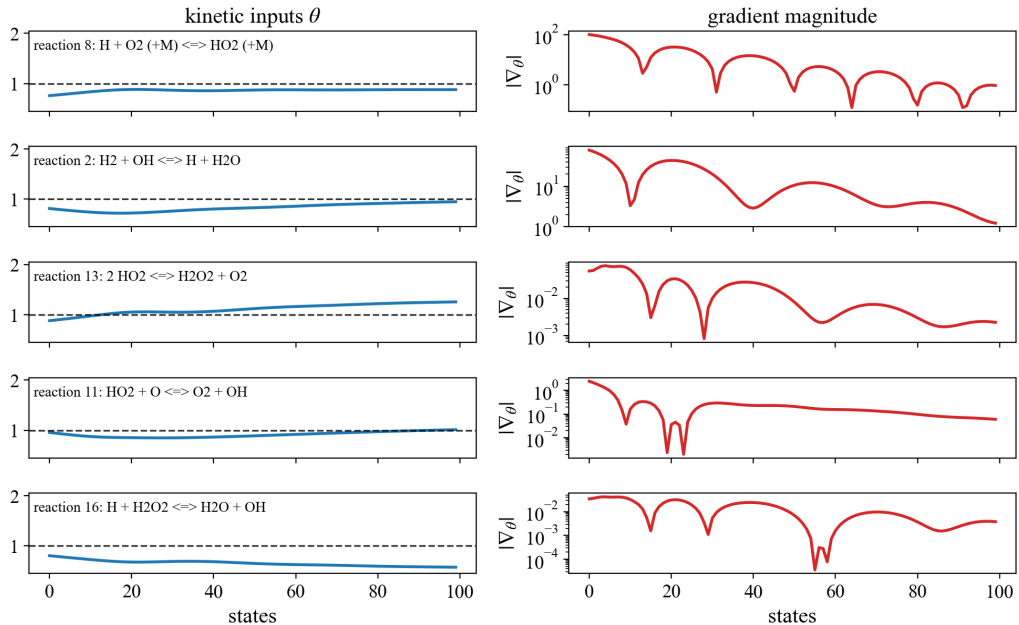

**Fig. 12.** Comparisons of specific reactions' kinetic inputs (left column) and gradients (right column) evolutions in Li-H2 BSS flame.

*3.4. Kinetics optimization with noise disturbance*

Although Diff-Chem Neural ODE can successfully achieve physically consistent optimization, practical experimental measurements often contain uncertainty and noise, which may propagate through the objective function and potentially bias the inferred kinetics. We therefore conduct a preliminary robustness study by injecting Gaussian measurement noise into the primal species used to construct the loss. Specifically, we perturb the primal species observations with random gaussian noise at levels of 1%, 5%, 10%, and 20%.

Fig. 13 summarizes the optimized loss for each case as a function of the noise level. Overall, the final loss does not exhibit a significant degradation as the noise increases. For hydrocarbon flames, the optimized loss is nearly unchanged, and in some cases even decreases slightly, which can be attributed to the stochastic nature of the perturbations and the fact that the optimization effectively averages noise across multiple locations/species in the objective. Hydrogen flames show a modest increase in the final loss with increasing noise, yet the optimization does not collapse; the loss remains within a low range, indicating stable convergence under noisy supervision. To quantify species-level accuracy, Fig. 14 reports the distribution of relative errors for two representative major species (e.g., $H_2$ and $O_2$) after optimization at different noise levels. The noise-induced performance degradation is moderate: the relative errors remain well controlled (within 2% for these major species), and the distribution shifts only slightly as the noise level increases. This suggests that the optimized solution is not overly sensitive to moderate measurement noise in the objective.

Furthermore, Fig. 15 and Fig. 16 visualize the optimized profiles for several representative species in the BSF and BSS flames under the 10% noise setting. In these figures, the gray dashed curves denote the noisy measurements used in the objective function, the dashed curves indicate the initial state, the solid curves represent the optimized state, and the black dash-dotted curves correspond to the clean reference data. Notably, even under noisy supervision, the optimized state aligns closely with the clean reference profiles, without exhibiting local, pointwise overfitting to the noisy targets. This behavior indicates that Diff-Chem Neural ODE provides an implicit regularization through its physics-consistent diffusion-chemistry coupling and the shared kinetic-parameterization across all sampled spatial locations. As a result, the optimizer is constrained to follow globally consistent kinetic updates that simultaneously explain multiple correlated observables, rather than fitting noise at isolated points. In other words, the proposed framework exhibits meaningful noise robustness: it can recover a physically consistent solution that generalizes beyond noisy measurements and remains faithful to the underlying clean flame dynamics.

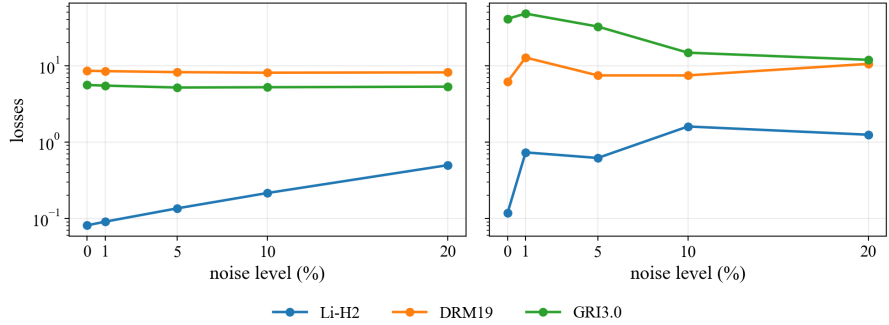

**Fig. 13. Optimized losses of BSF flames (left column) and BSS flames (right column) under different noise levels.**

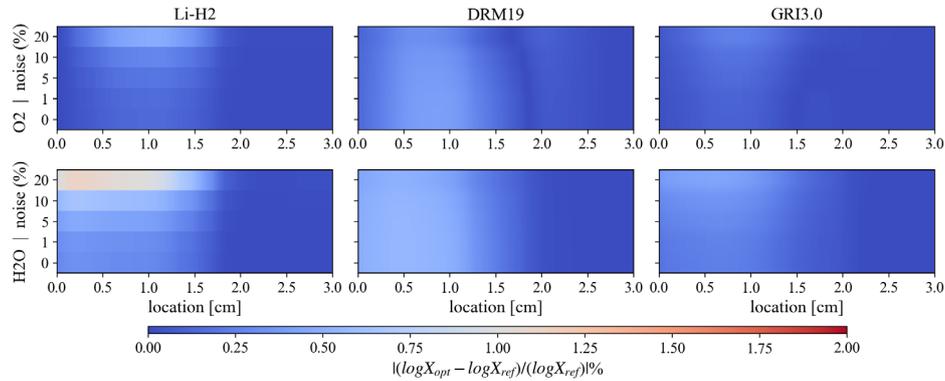

**Fig. 14. Optimized relative error of representative species in BSF flames under different noise levels.**

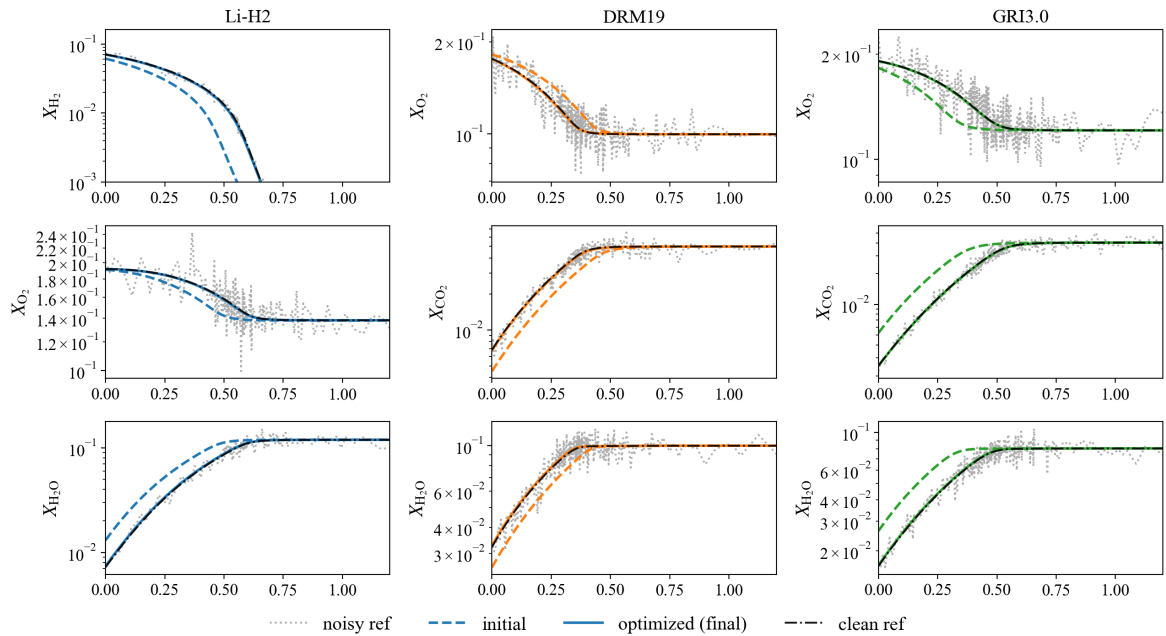

**Fig. 15. Comparisons of representative species of initial and optimized states in BSF flames under 10% noise level.**

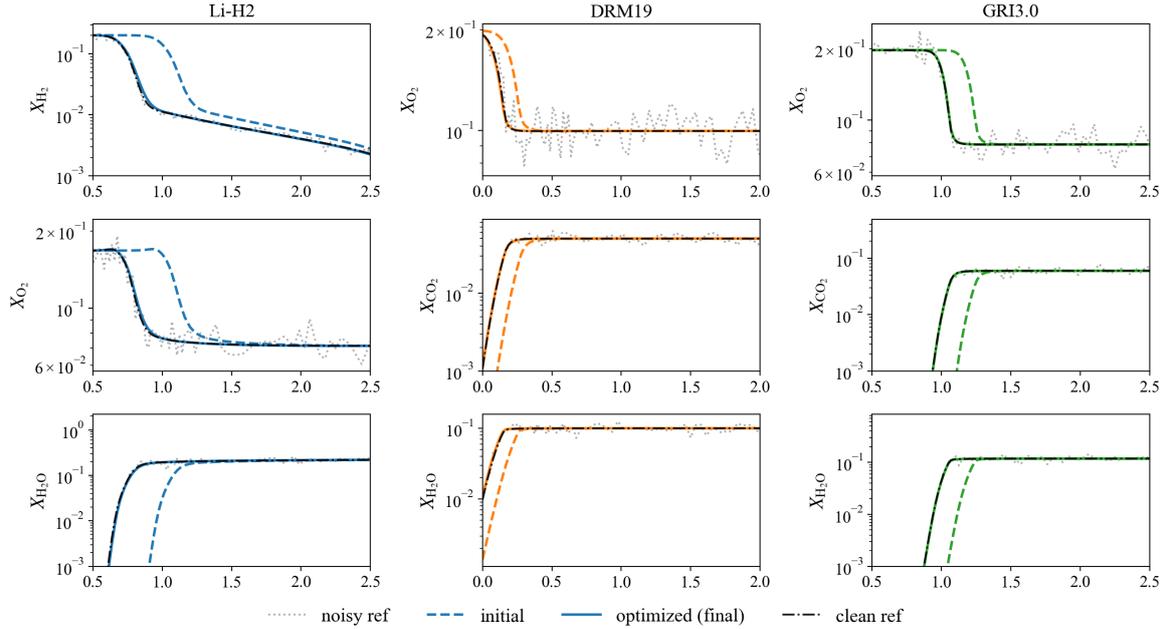

**Fig. 16. Comparisons of representative species of initial and optimized states in BSS flames under 10% noise level.**

## 4. Conclusion

In this work, we developed a transport-chemistry coupled neural ordinary differential equation framework for R-D system modeling and kinetics optimization. By embedding Arrhenius-structured reaction neurons into a fully differentiable architecture and explicitly accounting for diffusion coupling along the streamline formulation, the proposed method preserves physical consistency while remaining directly usable for gradient-based optimization without sampling-based pretraining. Comprehensive evaluations on BSF and BSS reacting flows, covering mechanisms with varying stiffness (Li-H2, DRM19, and GRI3.0), show that Diff-Chem Neural ODE reproduces full-PDE species profiles with near-reference accuracy. In contrast, a pure-chemistry Neural ODE that neglects transport coupling can introduce large deviations in species fields, misplace the ignition location, and yield unrealistically thin reaction zones, which in turn undermines mechanistic interpretation and sensitivity assessment near the flame front and in the pre-ignition region.

We further demonstrated that Diff-Chem Neural ODE is robust and effective for optimization applications under practical constraints. Under randomized initial-state perturbations, Diff-Chem Neural ODE consistently exhibits smaller and narrower error distributions than the pure-chemistry baseline. In gradient evaluation, it achieves substantial speedups compared to fully discretized PDE computations while maintaining the best accuracy and efficiency trade-off among the tested approaches when diffusion-chemistry coupling is essential. For kinetics refinement, using only a limited set of primal species in the objective, the method reliably reduces the loss by more than 98% and restores both flame structure and ignition location; notably, temperature and secondary species (which never directly included in the loss) also recover to the reference solution, indicating physically consistent global control of the coupled system. Sensitivity evolution further confirms that reactions with larger gradient magnitudes are preferentially corrected toward reference kinetic inputs, while low-gradient reactions exert limited influence, providing actionable guidance for prioritizing kinetic updates and revealing potential non-uniqueness in parameter identification. Finally, preliminary tests with measurement noise injected into primal species (1%-20%) show stable convergence without local overfitting: optimized profiles remain smooth and closely match clean references, with only moderate degradation in hydrogen cases and near-invariant performance in hydrocarbon flames.

Overall, Diff-Chem Neural ODE offers a scalable and physically faithful route to efficient kinetics optimization in diffusion-coupled reacting systems, particularly suited to settings with limited and noisy experimental observables. Future work will extend Diff-Chem Neural ODE to multi-dimensional flame configurations and broader transport models (e.g., differential diffusion and multicomponent transport) to further improve realism. We also plan to integrate explicit uncertainty quantification and noise-aware objectives to enable more reliable kinetic inference from sparse, heterogeneous experimental datasets.

## Declaration of competing interest

All authors declare that they have no conflict of interest to this work.

## Acknowledgments

This work was supported by grants from the National Natural Science Foundation of China (No. 52025062).